\documentclass[prl,letterpaper,twocolumn,nofootinbib,preprintnumbers,superscriptaddress]{revtex4-1} 

\pdfpageattr {/Group << /S /Transparency /I true /CS /DeviceRGB>>}

\usepackage[utf8]{inputenc}

\usepackage{titlesec}
\usepackage{dcolumn}
\usepackage{bm}
\usepackage{amsmath,amssymb,amsfonts}

\usepackage{colortbl}
\usepackage{xfrac}
\newcommand{\ra}[1]{\renewcommand{\arraystretch}{#1}}
\usepackage{booktabs}
\usepackage{graphicx}
\usepackage{color}
\usepackage{bbold}
\usepackage{lipsum}

\usepackage{slashed}
\usepackage[svgnames]{xcolor}
\usepackage[english]{babel}
\usepackage{blindtext}
\usepackage{microtype}
\usepackage{tikz}
\usepackage[most]{tcolorbox}
\usepackage{libertine}
\usepackage[colorlinks=true,linkcolor=blue,urlcolor=blue,citecolor=blue]{hyperref}
\usepackage{ifpdf}

\newtcbox{\mymath}[1][]{%
    nobeforeafter, math upper, tcbox raise base,
    enhanced, colframe=blue!30!black,
    colback=blue!30, boxrule=1pt,
    #1}

\usepackage{slashed}
\usepackage{selinput}
\SelectInputMappings{
    adieresis={ä},
    germandbls={ß}
}

\usepackage{libertine}

\newcommand{\Eq}[1]{Eq.~\eqref{#1}}

\newcommand{\Refs}[1]{Refs.~\cite{#1}}

\newcommand{\Fig}[1]{Fig.~{\ref{#1}}}

\newcommand{\Table}[1]{Table~\ref{#1}}

\renewcommand{\Ref}[1]{Ref.~\cite{#1}}

\newcommand{\vect}[1]{\boldsymbol{#1}}

\newcommand{\sh}[1]{\slashed{#1}}

\newcommand{\pslash}{\mbox{$\not \! p$}}
\newcommand{\aslash}{\mbox{$\not \! a$}}
\newcommand{\nslash}{\mbox{$\not \! n$}}

\newcommand{\qslash}{\mbox{$\not \! q$}}

\newcommand{\kslash}{\mbox{$\not \! k$}}
\newcommand{\vslash}{\mbox{$\not \! v$}}

\graphicspath{{.}{figures/}} 

\def\hs{\hspace}

\def\lf{\left}
\def\rg{\right}

\makeatletter
\newcommand*\if@single[3]{%
  \setbox0\hbox{${\mathaccent"0362{#1}}^H$}%
  \setbox2\hbox{${\mathaccent"0362{\kern0pt#1}}^H$}%
  \ifdim\ht0=\ht2 #3\else #2\fi
  }
\newcommand*\rel@kern[1]{\kern#1\dimexpr\macc@kerna}
\newcommand*\widebar[1]{\@ifnextchar^{{\wide@bar{#1}{0}}}{\wide@bar{#1}{1}}}
\newcommand*\wide@bar[2]{\if@single{#1}{\wide@bar@{#1}{#2}{1}}{\wide@bar@{#1}{#2}{2}}}
\newcommand*\wide@bar@[3]{%
  \begingroup
  \def\mathaccent##1##2{%
    \if#32 \let\macc@nucleus\first@char \fi
    \setbox\z@\hbox{$\macc@style{\macc@nucleus}_{}$}%
    \setbox\tw@\hbox{$\macc@style{\macc@nucleus}{}_{}$}%
    \dimen@\wd\tw@
    \advance\dimen@-\wd\z@
    \divide\dimen@ 3
    \@tempdima\wd\tw@
    \advance\@tempdima-\scriptspace
    \divide\@tempdima 10
    \advance\dimen@-\@tempdima
    \ifdim\dimen@>\z@ \dimen@0pt\fi
    \rel@kern{0.6}\kern-\dimen@
    \if#31
      \overline{\rel@kern{-0.6}\kern\dimen@\macc@nucleus\rel@kern{0.4}\kern\dimen@}%
      \advance\dimen@0.4\dimexpr\macc@kerna
      \let\final@kern#2%
      \ifdim\dimen@<\z@ \let\final@kern1\fi
      \if\final@kern1 \kern-\dimen@\fi
    \else
      \overline{\rel@kern{-0.6}\kern\dimen@#1}%
    \fi
  }%
  \macc@depth\@ne
  \let\math@bgroup\@empty \let\math@egroup\macc@set@skewchar
  \mathsurround\z@ \frozen@everymath{\mathgroup\macc@group\relax}%
  \macc@set@skewchar\relax
  \let\mathaccentV\macc@nested@a
  \if#31
    \macc@nested@a\relax111{#1}%
  \else
    \def\gobble@till@marker##1\endmarker{}%
    \futurelet\first@char\gobble@till@marker#1\endmarker
    \ifcat\noexpand\first@char A\else
      \def\first@char{}%
    \fi
    \macc@nested@a\relax111{\first@char}%
  \fi
  \endgroup
}
\makeatother

\begin{document}

\title{ Gluon PDF from Quark Dressing in the Nucleon and Pion}

\author{Adam Freese}
\affiliation{Physics Division, Argonne National Laboratory, Argonne, IL 60439 USA}

\affiliation{Department of Physics, University of Washington, Seattle, WA 98195 USA}

\author{Ian C. Clo\"et}
\affiliation{Physics Division, Argonne National Laboratory, Argonne, IL 60439 USA}

\author{Peter~C.~Tandy}
\affiliation{Center for Nuclear Research, Department of Physics, Kent State University, Kent OH 44242 USA}

\affiliation{CSSM, Department of Physics, University of Adelaide, Adelaide SA 5005, Australia}

\begin{abstract}
Gluon dressing of the light quarks within hadrons is very strong and extremely important in that it dynamically generates most of the observable mass through the breaking of chiral symmetry.   The quark and gluon parton densities, $q(x)$ and $g(x)$, are necessarily interrelated since  any gluon emission and absorption process, especially dressing of a quark, contributes to $g(x)$  and modifies $q(x)$. 
Guided by long-established results for the parton-in-parton distributions from a strict 1-loop perturbative analysis of a quark target, we extend the non-perturbative QCD approach based on the  Rainbow-Ladder truncation of the Dyson-Schwinger equations to describe the interrelated  valence $q_{\rm v}(x)$ and the dressing-gluon $g(x)$ for a hadron at its intrinsic model scale.   We employ the pion description from previous DSE work that accounted for the gluon-in-quark effect, and introduce a simple model of the nucleon for exploratory purposes.  We find typically \mbox{$\langle x \rangle_g \sim 0.20$}  for both pion and nucleon at the model scale, and the valence quark helicity contributes 52\% of nucleon spin.  We deduce both $q_{\rm v}(x)$ and $g(x)$ from 30 calculated Mellin moments, and after adopting existing data analysis results for $q_{\rm sea}(x)$, we find that  NLO scale evolution produces $g(x)$ in good agreement with existing data analysis results for the pion at 1.3~GeV 
and the nucleon at 5~GeV$^2$.    At the  scale 2~GeV typical of lattice-QCD calculations, we obtain \mbox{$\langle x \rangle_g^{\rm N} = 0.42$} in good agreement with 0.38 from the average of recent lattice-QCD calculations. 
\end{abstract}


\maketitle


\noindent\textbf{Introduction:} 
Recent progress in understanding the structure of hadrons is increasingly focussed on the separate roles of quarks and gluons.   A consensus is starting to emerge~\cite{Lin:2017snn,Lin:2020rut} from experimental and theoretical work on integral properties such as the quark and gluon parton contributions to the nucleon spin, angular momentum, and lightcone momentum~\cite{Deka:2013zha,Yang:2016plb,Yang:2018bft}.   
After a long time restricted to low moments of parton distribution functions (PDFs), the lattice-regulated approach to QCD calculations has in recent years developed methods to obtain the momentum fraction $x$-dependence of PDFs, see e.g., \Refs{Sufian:2019bol,Gao:2020ito}.

Gluons have two very strong dynamical roles in hadron physics.  Besides their role in binding quarks, perhaps the next most prominent role is the generation of around 95\% of the mass of most light quark hadrons through the mechanism of dynamical breaking of chiral symmetry (DCSB).  It is possible that one of these two roles dominates the gluon parton structure of such hadrons.  Lattice-QCD calculations of the gluon fraction of the nucleon spin and lightcone momentum are typically 35-50\% at scale 2~GeV~\cite{Yang:2016plb,Yang:2018bft,Shanahan:2018jbm}.   Any gluon emission and absorption process,  including dressing of a quark, contributes to the gluon parton density $g(x)$  and modifies the quark part $q(x)$~\cite{Collins:2011zzd}.    
The concept of a PDF of a parton in a parton is often used in considerations of radiative processes that describe changes with resolving scale~\cite{Korchemsky:1988si,Berger:2002sv}, pQCD issues within factorization~\cite{Ji:2004hz,Liu:2020rvc}, and explorations of the relation between 1-loop quark dressing and the consequent gluon-in-quark and quark-in-quark PDFs~\cite{Collins:2011zzd,Bringewatt:2020ixn}.    The latter work illustrates that  a gluon dressing mechanism produces both \mbox{$\langle x \rangle_q < 1$}  and \mbox{$\langle x \rangle_g > 0$} at any scale.  It has recently been found that an assumption of zero gluon and sea distributions at low scales typical of models is incompatible with scale evolution and current data~\cite{Diehl:2019fsz}.  
The application of the canonical QCD definition~\cite{Collins:1981uw} of $g(x)$ (\Eq{eq:GPDF_defn} below) to obtain the gluon-in-quark $g(x)$ at the intrinsic scale of a non-perturbative hadronic model containing  DCSB has not been made before. 

We investigate this within the Dyson-Schwinger equation approach (DSE) to hadron physics which employs the ladder-rainbow truncation of diagrams; an infinite subset of gluon emission and absorption processes are thereby included.
This DSE-RL approach has proven  to be very efficient for ground state masses, decay constants, and electromagnetic form factors~\cite{Bashir:2012fs,Cloet:2013jya,Tandy:2014hja,Horn:2016rip}.  It has been  especially accurate for light quark pseudoscalar and vector mesons~\cite{Maris:1999nt,Maris:2000sk} because their properties are strongly dictated by the dynamical breaking of chiral symmetry and vector current conservation, which is built into the approach.  It has been applied to pion, kaon and nucleon PDFs~\cite{Nguyen:2011jy,Chang:2014lva,Chang:2014gga,Chen:2016sno,Bednar:2018htv,Shi:2018mcb,Ding:2019lwe} mostly using the Ward Identity Ansatz to represent the relevant quark vertex for any PDF moment.  As discussed later, this approximation is accurate only for the lowest (quark number) moment, and does not distinguish the gluon-in-quark and quark-in-quark PDFs.   One DSE approach that does make this distinction has been applied to $q(x)$ for the pion~\cite{Bednar:2018mtf}.  

\medskip
\noindent\textbf{Quark and Gluon PDFs in the Nucleon:} 
In the Bjorken kinematical limit, and at leading-twist, the unpolarized quark and gluon parton densities  are defined by the explicitly Poincar\'e-invariant matrix elements~\cite{Collins:1981uw,Jaffe:1983hp,Jaffe:1996zw,Diehl:2003ny} 
\begin{align}
q(x) = \int \frac{d\lambda}{4\,\pi}\  e^{-i x P\cdot n \lambda} \,
\left< P\left|\bar{q}(\lambda n)\,\sh{n}\,W(\lambda,A)\,q(0)\right|P\right>_c~,
\label{eq:PDF_defn}
\end{align}
and 
\begin{align}
g(x) = \int \frac{d\lambda}{2\,\pi} \frac{e^{-i x P\cdot n \lambda} }{x P^+} \,
\left< P\left| G^{+\, \mu}(\lambda n)\,W\,G_{\mu \,+}(0) \right|P\right>_c~.
\label{eq:GPDF_defn}
\end{align}
Here $n^\mu$ is  the  light-like longitudinal basis vector (given by $(1,\,\vect{0}_T,\,-1)$ in the target rest frame),   $W(\lambda, A)$ is the  Wilson line integral that restores  gauge invariance to the non-local current, and
\mbox{$G^{+\, \mu}(\lambda n) = n_\nu \,G^{\nu \mu}(\lambda n)$}.
At the same level, the helicity gluon PDF is given by 
\begin{align}
\Delta g(x) = i \int \frac{d\lambda}{2\,\pi} \frac{e^{-i x P\cdot n \lambda} }{x P^+} \,  
\left< P\left| G^{+\, \mu}(\lambda n)\,W\,\tilde{G}_{\mu \,+}(0) \right|P\right>_c
\label{eq:polGPDF_defn}
\end{align}
where the dual tensor is \mbox{$\tilde{G}^{\mu \nu}(z) = \frac{1}{2}\epsilon^{\mu \nu \alpha \beta} \,G_{\alpha \beta}(z)$}.
The quark helicity PDF is given by \Eq{eq:PDF_defn} with the replacement \mbox{$\sh{n} \to \sh{n}\, \gamma_5$}.

As pointed out 
some time ago~\cite{Collins:1981uw} parton momentum conservation is a consequence of the above formal definitions, and is gauge-invariant, and scale-invariant. 
\begin{figure}[tbp]
\vspace*{-3mm}\centering\includegraphics[width=0.8\columnwidth,height=84mm]{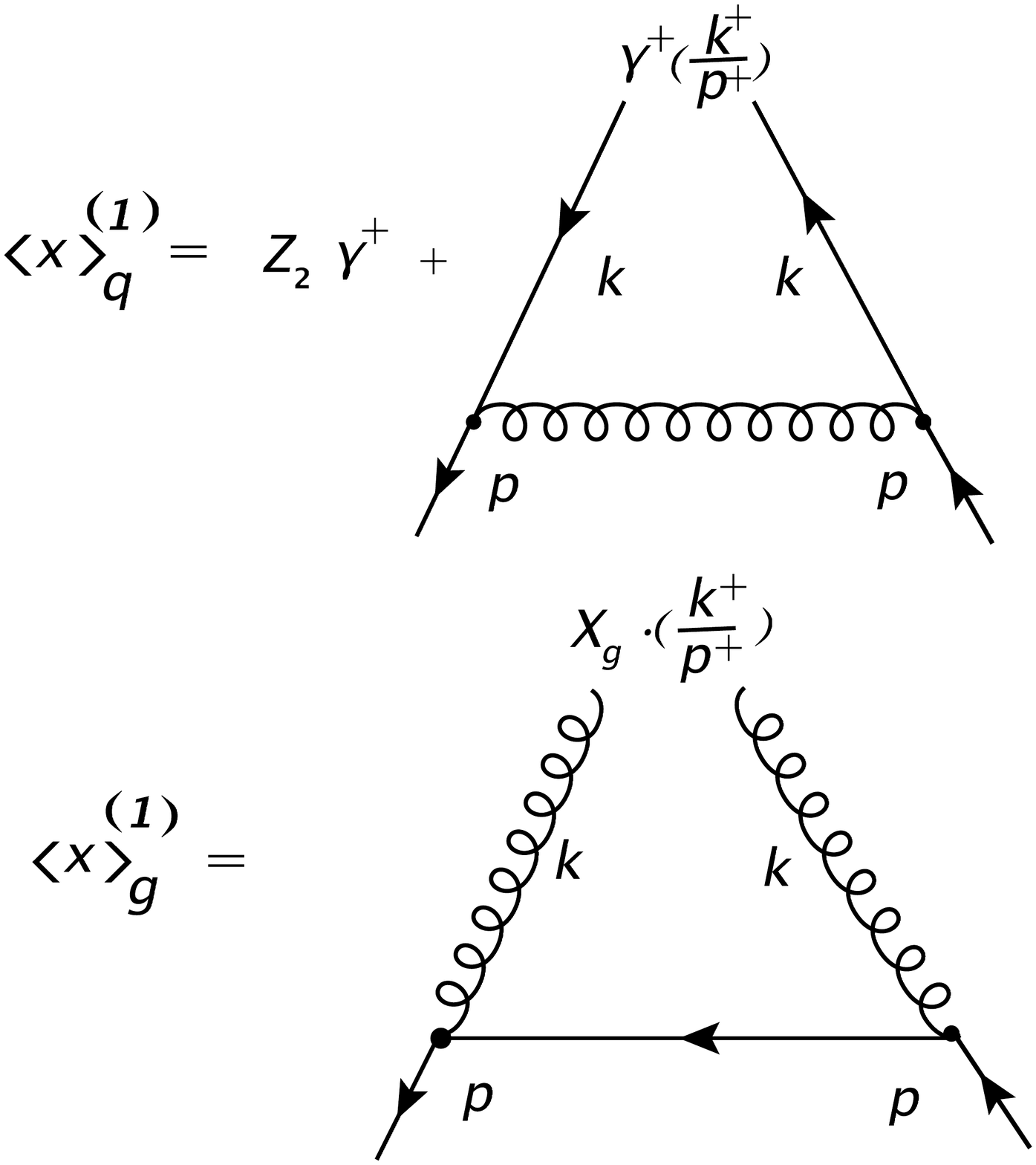} 
\vspace*{-7mm}\caption{Diagrammatic illustrations of the PDF momentum fractions for a quark target treated at 1-loop order.  The top part illustrates \protect\Eq{eq:Quark_1loop_LC}  for $\langle x \rangle_q$; the bottom part represents  \protect\Eq{eq:Gluon_1loop} for $\langle x \rangle_g$. } 
\label{fig:1-loop_Diags}  
\end{figure}
The sum of the unpolarized PDF momentum fractions 
from  the above expressions gives 
\begin{align}
\mathcal{S}_x = \small\sum_q\, \langle x \rangle_q + \langle x \rangle_g =  \frac{1}{2 (P^{+})^2}&  \left< P\left| \bar{q}(0)\, \right. \right.  i\sh{n}\, D^{+}\, q(0) 
\nonumber \\
&\hs*{-6mm}\left. \left.  +G^{+\, \mu}(0)\,G_{\mu \,+}(0) \right|P\right>_c,
\label{eq:MomSum}
\end{align}
where \mbox{$ D^{+} = n_\nu D^\nu $} is the lightcone projected covariant derivative, and
$\small\sum_q$ includes antiquarks and all relevant flavors.  The operator density on the RHS of \Eq{eq:MomSum} is proportional to the light-cone projection of the energy-momentum tensor density \mbox{$T^{+ +}(x) = n_\mu \,T^{\mu \nu}(x) \, n_\nu $}, where $T^{\mu \nu}(x)$ has the Belinfante improvement and is symmetric and gauge invariant~\cite{Ji:1995sv}.   The sum rule \mbox{$ \mathcal{S}_x =  $}  \mbox{$ \left< P\left| T^{+ +}(0) \right|P\right>_c/2 (P^+)^2 = 1$}
then follows after covariant normalization  \mbox{$ \left< P^\prime | P \right>  = 2 E \,\delta^3(P^\prime - P) $}.    
\begin{figure}[tbp]
\vspace*{-18mm}\centering\includegraphics[width=0.6\columnwidth,height=78mm]{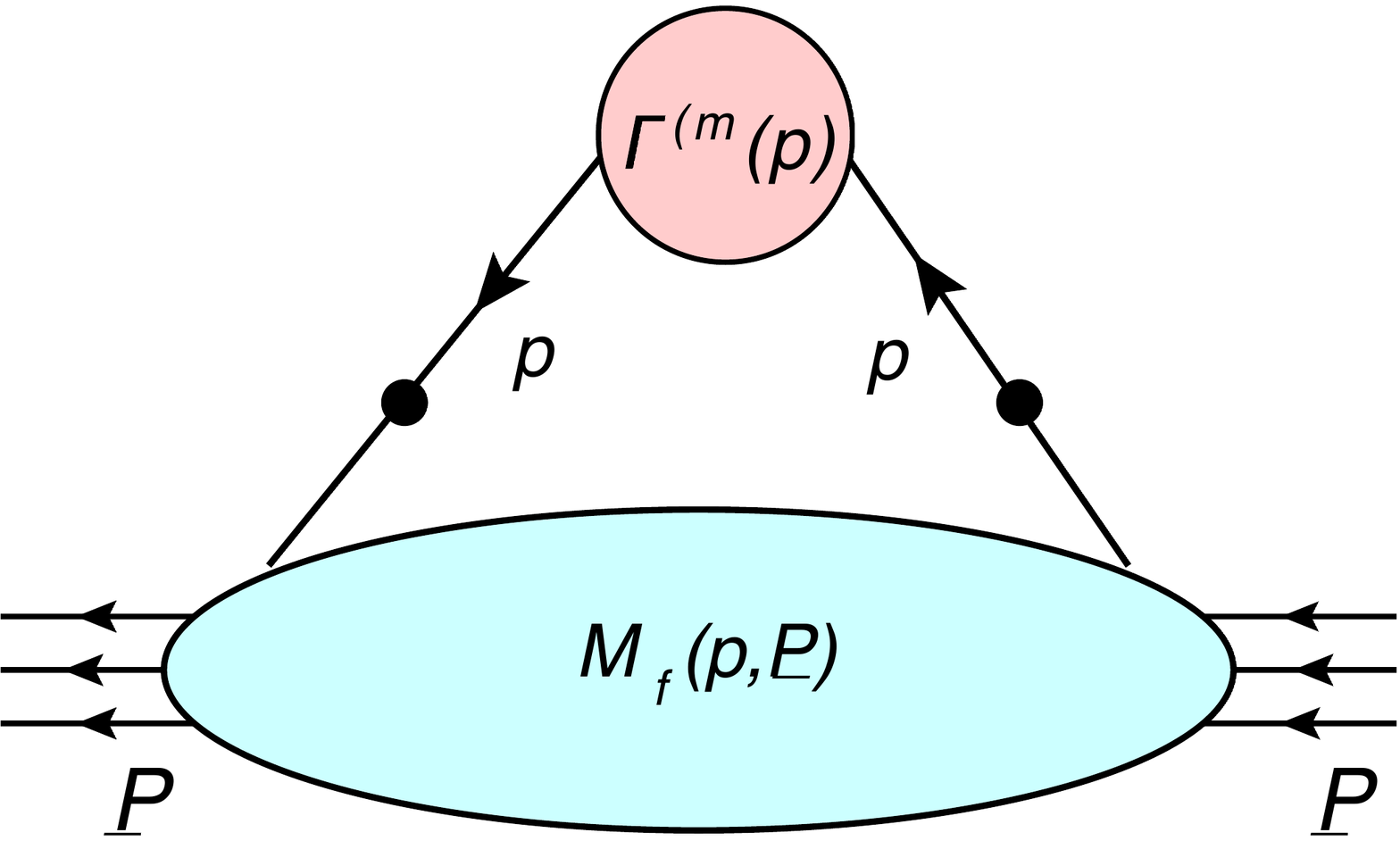}\\[0.5em]
\vspace*{-62mm}\centering\includegraphics[width=0.94\columnwidth,height=126mm]{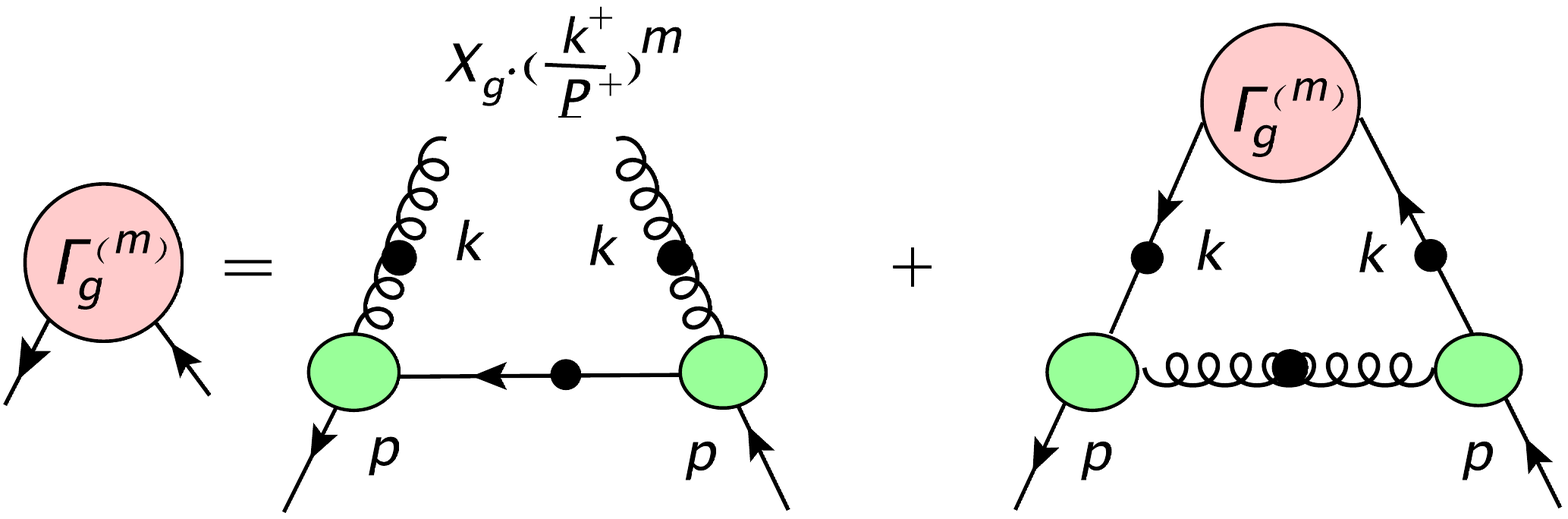}
\vspace*{-44mm}\caption{{\it Top panel:}  An illustration of the loop integral given in \protect\Eq{eq:gen_q_moms_N} for 
$\langle x^m \rangle_q$  or  in \protect\Eq{eq:gen_g_moms_N} for $\langle x^m \rangle_g$ using BSE-generated vertices as appropriate to each case.
{\it Bottom panel:}  An illustration of the BSE integral equation \protect\Eq{eq:Gamma_BSE} for the  vertex that carries the moment information of the PDF of the gluon-in-quark, as specified by the particular  inhomogeneous term given in \protect\Eq{eq:Inhom_GVertm}.    The latter is the dressed generalization of the 1-loop counterpart in the bottom of \protect\Fig{fig:1-loop_Diags}. 
}
\label{fig:Gen_Triangle_Diag}  
\end{figure}

\medskip
\noindent\textbf{Quark target at 1-loop:} 
We outline relevant partonic aspects of a single quark "hadronic" target, treated at 1-loop some time ago~\cite{Collins:2011zzd}.  The previous definitions employed Euclidean metric; from here on we adopt Euclidean metric as a prelude to the realistic numerical treatment of  non-perturbative aspects using Rainbow-Ladder truncation which is  defined and well established in Euclidean metric.\footnote{In Euclidean metric we employ $a_4 = i a^0$ for any space-time vector, including $n$, while  \mbox{$\{\gamma_\alpha, \, \gamma_\beta\} =\delta_{\alpha \beta} $}  with $\gamma_4 = \gamma^0$.   Hence \mbox{$\aslash \to -i \aslash $} while \mbox{$a \cdot b \to - a \cdot b $}.  }  From \Eq{eq:PDF_defn},  the 1-loop result for $\langle x \rangle_q$, as displayed in \Fig{fig:1-loop_Diags}, is given by 
\begin{align}
\langle x \rangle_q^{(1)} = {\rm tr_d} \;\bar{u}_{1} \,\Big[ Z_2\,(-i \nslash ) - \frac{4 g^2}{3}  \int_k^R \frac{k\cdot n}{p\cdot n } \, \frac{T_q(p,k)}{q^2} \Big] \,u_{1}~, 
\label{eq:Quark_1loop_LC}
\end{align}
where $p$ is the quark target momentum,  $x=k\cdot n/p\cdot n$,  and for simplicity we have employed light-cone (LC) gauge to eliminate the Wilson line.  Here
\begin{align}
T_q(p,k) = \gamma_\mu\, V \, \gamma_\mu  - \frac{\qslash \,V \,\nslash + \nslash \,V \,\qslash }{q\cdot n} 
\label{eq:Triag_q_LC}
\end{align}
with  \mbox{$V = S(k)\,(-i \nslash ) \, S(k) $}  and gluon momentum $q=p-k$.  The spinor product is for a fully polarized quark  and the normalization is such that \mbox{$\bar{u}_1 \,u_1  = 1$} \mbox{$= \bar{u}_1 \, (-i\nslash )  \,u_1 $}.  The notation $\int_k^R$ represents $\int^R \frac{d^4k}{(2 \pi)^4}$ with $R$ indicating a smooth regularization in both the deep ultraviolet and deep infrared.  The gauge-invariant quark helicity $\Delta_q $ is obtained from \Eq{eq:Triag_q_LC} by replacement of $(-i \nslash )$ by $(-i \nslash \, \gamma_5)$ throughout, and setting \mbox{$k\cdot n/p\cdot n \to 1$}.

From  \Eq{eq:GPDF_defn}, the Wilson line makes no contribution to $\langle x \rangle_g$.    At 1-loop, the gluon field can only arise from dressing of the quark and, as displayed in \Fig{fig:1-loop_Diags},  the gauge-invariant 1-loop result  is
\begin{align}
\langle x \rangle_g^{(1)} = \bar{u}_{1} \,\Big[ \frac{4 g^2}{3}  \int_k^R \frac{k\cdot n}{p\cdot n } \frac{2\, \hat{X}_g(p,k) }{k^4}  \Big] \,u_{1}~, 
\label{eq:Gluon_1loop}
\end{align}
where again $x=k\cdot n/p\cdot n$.   Here  
\begin{align}
\hat{X}_g(p,k) =  k\cdot n \,\Big( \gamma_\mu \,S(q) \,\gamma_\mu - &\frac{ \nslash \,S(q) \,\kslash + \kslash \,S(q) \,\nslash }{k\cdot n}  \nonumber \\
& + \frac{\nslash \, k^2\, S(q) \, \nslash }{(k\cdot n)^2} \Big)~,
\label{eq:Inhom_Xg}
\end{align}
and \mbox{$q = p-k$} is the loop quark momentum.    Integration by parts applied to the above confirms that  \mbox{$\langle x \rangle_q^{(1)} = $} \mbox{$\bar{u}_1 [- n\cdot \partial_p\, S^{-1}(p) ] u_1 - \langle x \rangle_g^{(1)} $} where the first term involves the Ward Identity vertex in terms of  the 1-loop propagator $S(p)$.   Due to the conserved vector current, the first (Ward Identity vertex) term is the unit quark number (thus fixing $Z_2$) and the momentum sum is verified.  The gauge-invariant 1-loop gluon helicity is given by 
\begin{align}
\Delta g^{(1)} =  \bar{u}_{1}\, \Big[ - 2\,\frac{4\, g^2}{3} \int_k^R \,\epsilon^{\mu + \alpha \beta} \, \frac{k_\alpha \, \gamma_\mu \,S(q)\, \gamma_\beta}{k^4} \Big] \, u_{1} ~ .
\label{eq:DeltaG_1loop}
\end{align}

Typical 1-loop results are  
\begin{align}
\begin{tabular}{c|cc|cc} 
              \hspace{1mm} $g^2$ \hspace{1mm} &   \hspace{2mm} $\langle x \rangle_q^{(1)} $  & \hspace{3mm}   $ \langle x \rangle_g^{(1)} $\hspace{3mm}   & \hspace{2mm} $\Delta_q^{(1)} $ \hspace{2mm}&  \hspace{2mm} $2 \Delta g^{(1)}$                     \\
\hline
                       5.78       &     0.766   & 0.234      &   0.974      &    0.203          \\
\end{tabular}~~~,
\label{eq:1loop_results}
\end{align}
where $g^2$ has been chosen to yield a magnitude for $ \langle x \rangle_g^{(1)} $ typical of the non-perturbative results for the pion and nucleon to be discussed below.   The integrals are regularized via the proper time method with \mbox{$\Lambda_{\rm IR} = 0.05~{\rm GeV}$} and \mbox{$\Lambda_{\rm UV} = 20~{\rm GeV}$}.    
The above quark target expressions are the 1-loop limit of the \mbox{$m=1$} PDF moments given by  \mbox{$\langle x^m \rangle = $}  \mbox{$\bar{u}_1\,  \Gamma^{(m)}(p) \, u_1$} in terms of the appropriate dressed  quark vertex that carrying information on the quark-in-quark or the gluon-in-quark PDF.  This example illustrates that any gluon radiation, absorption or splitting dynamics, especially dressing of a quark, generates linked contributions to both $q(x)$ and $g(x)$.   

\medskip
\noindent\textbf{Hadron PDFs:} 
To generalize the above 1-loop quark vertex structures and apply them to valence quarks as offered by a hadron,
we employ the Rainbow-Ladder truncation of the DSE approach that has successfully described many hadron properties~\cite{Bashir:2012fs,Cloet:2013jya,Tandy:2014hja,Horn:2016rip}.  Without the Wilson line contribution, \Eq{eq:PDF_defn} applied to the nucleon in this DSE-RL approach produces 
\begin{align}
q_f(x) = {\rm tr} \int_p^\Lambda  S(p)\, \Gamma_q(p,x)\, S(p)\,{\mathcal M}_f(p,P) \, , 
\label{eq:Gammax_N}
\end{align}
where  the trace is over Dirac and color indices, and $\int_p^\Lambda$ represents $\int^\Lambda \frac{d^4p}{(2 \pi)^4}$ with $\Lambda$ indicating the ultraviolet regularization mass scale. 
The quark vertex $\Gamma_q(p,x)$ is generated from the inhomogeneous term $ Z_2 \, (-i \nslash) \,\delta(x - p \cdot n/P \cdot n)$ via the  Bethe-Salpeter integral equation~\cite{Nguyen:2011jy,Bednar:2018mtf}.  
The nucleon amplitude $ {\mathcal M}_f(p,P) $ has the Dirac spinor structure of a  $\bar q$-nucleon scattering amplitude, 
and describes the probability amplitude for the target to present a dressed quark of flavor $f$ and momentum $p$ that in turn yields a quark-in-quark with momentum fraction $x$  to the hard DIS probe.   The moments of the corresponding unpolarized quark PDFs are then
\begin{align}
\langle x^m \rangle_{q_f} = {\rm tr} \int_p^\Lambda  S(p)\, \Gamma_q^{(m)}(p)\, S(p)\,{\mathcal M}_f(p,P) \,, 
\label{eq:gen_q_moms_N}
\end{align}
where \mbox{$ \Gamma_q^{(m)}(p) = \int_0^1 dx \,x^m \, \Gamma_q(p,x)$}.   This nucleon case is illustrated in \Fig{fig:Gen_Triangle_Diag}.  For a pion $ {\mathcal M}_f(p,P) $ is replaced by $\bar{\Gamma}_\pi \, S(P-p)\, \Gamma_\pi  $ in terms of Bethe-Salpeter vertices $\Gamma_\pi $, see \Ref{Bednar:2018mtf}.  Our main emphasis is the nucleon and the present exploratory model for $ {\mathcal M}_f(p,P) $ is explained later.  The 1-loop version of  the vertex $\Gamma_q^{(m)}(p)$ in \Eq{eq:gen_q_moms_N} has been used in the quark target example discussed earlier.
\begin{table}[tbp]
\ra{0.9}

\begin{tabular}{c|c|ccc}\hline
                  & $\mu~(GeV)$  & \hspace{2mm} $2\,\langle x \rangle_{u_{\rm v}} $ \hspace{1mm}   &  \hspace{1mm} $ \langle x \rangle_{\rm sea} $  & \hspace{2mm} $\langle x \rangle_{\rm g} $ \hspace{2mm}    \\
\hline
\rule{0em}{3ex}    
    $\pi$: \hspace*{4mm} Here  \hspace*{4mm}       & 0.78    & 0.646    & 0.151  & 0.203            \\   
\hline
\rule{0em}{3ex}
    $\pi$: \hspace*{4mm} Here  \hspace*{4mm}      & $1.3$    & 0.540                & 0.156                  & 0.304    \\   
    \hspace*{5mm}JAM~\cite{Barry:2018ort}        & $1.3$    & $0.54 \pm 0.01$    & $0.16 \pm 0.02$    & $0.30 \pm 0.02$      \\   
\hline
\rule{0em}{3ex}    
    $\pi$: \hspace*{4mm} Here  \hspace*{4mm}     & $\sqrt{10}$    & 0.448                  &  0.174                    & 0.378    \\        
    \hspace*{5mm}JAM~\cite{Barry:2018ort}       & $\sqrt{10}$   & $0.45 \pm 0.01$  &  $0.17 \pm 0.01$   & $0.37 \pm 0.02$        \\   
%
\hline 
\end{tabular}
\caption{Momentum fractions from the $\pi^+$ PDFs at the model scale and at selected higher scales compared to global data analysis by the JAM collaboration. }
\label{Pion_m1_mu0_mu}
\end{table}
%

This Bethe-Salpeter equation for the  vertex is 
\begin{align}
\Gamma^{(m)}(p) =&  \Gamma^{(m)}_{D} (p) \nonumber \\
& - \int_\ell^\Lambda \frac{\lambda^a}{2} \gamma_\mu \mathcal{K}_{\mu \nu}(p-\ell)\, S(\ell)\, \Gamma^{(m)}(\ell)\, S(\ell)\, \frac{\lambda^a}{2} \gamma_\nu~, 
\label{eq:Gamma_BSE}
\end{align}
where the DSE-RL kernel $\mathcal{K}_{\mu \nu}(q)$ is given in \Eq{eq:RL_Kernel} below, and the inhomogeneous term for the unpolarized quark PDFs is \mbox{$\Gamma^{(m)}_{D} (p) =  $} \mbox{$Z_2\,(-i \nslash)\,(\frac{p\cdot n}{P \cdot n})^m$}.  The vertex appropriate to the quark helicity moments $\langle x^m \rangle_{\Delta q_f}$ is obtained by the substitution \mbox{$ (-i \nslash) \to (-i \nslash \gamma_5)$} within the inhomogeneous term.   

The moments of the unpolarized dressing gluon PDFs are  
\begin{align}
\langle x^m \rangle_{g_f} = {\rm tr} \int_p^\Lambda  S(p)\, \Gamma_g^{(m)}(p)\, S(p)\,{\mathcal M}_f(p,P) \,, 
\label{eq:gen_g_moms_N}
\end{align}
where  the 1-loop version of vertex $\Gamma_g^{(m)}(p)$ is illustrated in the previous section.  The DSE-RL version is given by solution of \Eq{eq:Gamma_BSE} with inhomogeneous term 
\begin{align}
\Gamma^{(m)}_{g,D} (p) = \frac{4}{3} \int_k^\Lambda \,\big(\frac{k\cdot n}{P \cdot n}\big)^m \, 2\,\hat{X}_g(p,k) \, {\mathcal K}_g(k^2) \, ,
\label{eq:Inhom_GVertm}
\end{align}
where $\hat{X}_g(p,k)$ is given by \Eq{eq:Inhom_Xg} except here the quark propagator is dressed.
The corresponding helicity gluon PDF moments use the quark vertex $\Gamma_{\Delta g}^{(m)}(p)$ which is generated from the BSE with the inhomogeneous term 
\begin{align}
\Gamma^{(m)}_{\Delta g,D} (p) = -\frac{4}{3} \int_k^\Lambda \big(\frac{k\cdot n}{P \cdot n}\big)^m \,2\,\epsilon^{\mu + \alpha \beta} \, k_\alpha \, \gamma_\mu \,S(q)\, \gamma_\beta \, {\mathcal K}_g(k^2) \, ,
\label{eq:Inhom_DelGVertm}
\end{align}
with \mbox{$ q = p - k$}.   The combination $g^2/k^4$ of the 1-loop formulas  has been generalized to ${\mathcal K}_g(k^2)$ to account for the non-perturbative dressing.    Details are given below and in the Appendix.  After the BSE is solved for $\Gamma_{\Delta g}^{(m)}(p)$ it is found that the inhomogeneous term \Eq{eq:Inhom_DelGVertm} is an excellent numerical approximation to the solution at the level of $10^{-3}$. 

\medskip
\noindent\textbf{Interaction Kernels:} 
In all cases, the dressed quark propagator $S(p)$ is obtained as the solution of QCD's quark Dyson-Schwinger equation in Rainbow-Ladder truncation, which is 
\begin{align}
S^{-1}(p) = Z_2\,S_0^{-1}(p) - 
\int_k^\Lambda \frac{\lambda^a}{2} \gamma_\mu \,\mathcal{K}_{\mu \nu}(q)\,  S(k)\, \frac{\lambda^a}{2} \gamma_\nu~,
\label{SigmaRL}
\end{align}
where \mbox{$ S_0^{-1}(p) = i \slashed{p} + Z_m m_r $}, $m_r$ is the renormalized current quark mass. The general form of the solution is 
$S^{-1}(p)  = i \pslash A(p^2, \zeta^2) + B (p^2, \zeta^2)$, where $\zeta$ is the renormalization scale where \mbox{$A \to 1$} and \mbox{$B \to m_r$}.  
The  standard DSE-RL interaction kernel~\cite{Maris:1999nt,Nguyen:2011jy,Qin:2011dd} that generates quark propagators and BSE vertices and meson bound states is \mbox{$ \mathcal{K}_{\mu \nu}(q) = {\mathcal K}_{\rm RL}(q^2) \,D^{\rm free}_{\mu \nu} (q)$} where
\begin{align}
{\mathcal K}_{\rm RL}(q^2) = D_{\rm RL}^2  \, {\rm e}^{-q^2/\omega^2} 
+ {\mathcal F}(q^2)\, 4\pi\, \tilde{\alpha}_s(q^2)  \, .
\label{eq:RL_Kernel}
\end{align}
Here $\tilde{\alpha}_s(q^2)$ denotes a continuation of the 1-loop $\alpha_s(q^2)$ to provide smooth non-singular coverage for the entire domain of $q^2$.   The first term of \Eq{eq:RL_Kernel} implements the infrared enhancement due to dressing effects, while the second term, with \mbox{${\cal F}(q^2)=$} \mbox{$(1 - \exp( -q^2/(1~{\rm GeV^2}) ) )/q^2 $}, connects smoothly with the 1-loop renormalization group behavior of QCD.  The DSE-RL kernel correlates a large amount of hadron physics~\cite{Bashir:2012fs,Cloet:2013jya,Tandy:2014hja,Horn:2016rip}.   

For the vertex that generates the gluon PDF,  the kernel ${\mathcal K}_g(k^2)$ can be identified by a generalization of the procedure that defines the standard BSE-RL kernel ${\mathcal K}_{\rm RL}(k^2)$ for the Bethe-Salpeter meson bound state equation which is linked by global symmetries to the dressed quark Dyson-Schwinger equation.  For example see \Ref{Maris:1997tm}.    Here we are using the properties of multiplicative renormalizability to use the large  renormalization scale dependence of propagators and vertex functions to produce their ultraviolet momentum dependence.   The BSE-RL kernel for the interaction of 2 quark currents collects the ultraviolet 1-loop momentum dependence from \mbox{$Z_2^2\, Z_3/Z_{1F}^2 \to $}  \mbox{$4 \pi\,\alpha_s(q^2) \to 4 \pi^2 \, \gamma_m /{\rm ln}(q^2/\Lambda_{\rm QCD}^2)$},  where \mbox{$ \gamma_m = 4/\beta_0 = 12/(33-2\, N_f)$}.   In the present case there is an extra dressed gluon propagator, and hence the deep ultraviolet behavior is characterized by an extra factor $Z_3(q^2,\Lambda^2)/q^2$. 

We take the interaction kernel ${\mathcal K}_g(q^2)$ to be the related form
\begin{align}
{\mathcal K}_g(q^2) = D_{\rm gg}^4  \, {\rm e}^{-q^2/\omega_g^2}
+ {\mathcal F}^2(q^2)\, 4\pi\, \tilde{\alpha}_s(q^2) \,\tilde{Z}_3(q^2) \, ,
\label{eq:CutG_Kernel}
\end{align}
where  $\tilde{Z}_3(q^2)$ denotes a continuation of the corresponding 1-loop $q^2$ dependence after the regularization mass scale $\Lambda$ has been absorbed into the definition of scale $\Lambda_{\rm QCD}$.    Details and parameters are given in the Appendix.

\medskip
\noindent\textbf{Model Nucleon Amplitude:} 
Firstly consider a collection of 3 non-interacting quarks, each with the same momentum $p_c=P/3$ and mass \mbox{$M_c = M_{\rm N}/3$},  and with spin-flavor probabilities ${\mathcal P}_{s,f}$.  The net flavor numbers \mbox{$n_f = \small\sum_s {\mathcal P}_{s,f} $} and polarizations \mbox{$ {\mathcal P}_f= $} \mbox{$ \small\sum_s {\mathcal P}_{s,f}\; s $} can be formally expressed as the matrix elements \mbox{$ {\rm tr}\,[ D_f(p_c)\; ( -i\nslash )] $} \mbox{$= n_f$} and \mbox{${\rm tr}\,[ D_f(p_c)\; ( -i\nslash \, \gamma_5 )] $} \mbox{$={\mathcal P} _f$}, where
\begin{align}
D_f(p_c) = \small\sum_s {\mathcal P}_{s,f}\, u_s\, \bar{u}_s =  \frac{-i\,\pslash_c + M_c}{2\,M_c}\,\big( \frac{n_f -i\,\gamma_5\,\vslash_f}{2}\, \big)~.
\label{eq:Duubar_f}
\end{align}
Here 
\mbox{$v_f $} \mbox{$ =(0, {\mathcal P}_f ; \vec{0} )$} is the polarization 4-vector.   The inclusion of interaction effects from gluon exchange and the extension to PDF moments for any $m$ are accomplished by replacement of  bare vertices such as $(-i\nslash) $  and $ (-i\nslash \, \gamma_5)$ by their dressed vertex counterparts $ \Gamma^{(m)}(p_c)$.  This connects with the 1-loop quark target case discussed earlier.  Similar comments apply for quark and gluon polarizations.

\begin{table}[tbp]
\addtolength{\tabcolsep}{1.7pt}
\addtolength{\extrarowheight}{1.0pt}
\begin{tabular}{l|cccc}
\hline\hline
PDF  & $N$ & $\beta$ & $\lf[a, b, c,d\rg]$  & $h$ \\
\hline
$g_N(x)$                  & $1.152$   & $-0.617 $      & $\lf[-2.80,\,6.465,\,2.144,\,0.0801\rg]$  & 4 \\
$g_\pi(x)$                 & $2.136 $  & $-0.663 $      & $\lf[2.094,\,-0.060,\,1.417,\,0.0122\rg]$  & 3 \\
\hline
\end{tabular}
\caption{ Fit parameters employed in \Eq{eq:BernsteinB4} for $g(x)$  at model scale $\mu_0$ for the pion ($\mu_0 = 0.78\,$GeV)  and for the nucleon ($\mu_0 = 0.56\,$GeV).
}
\label{tab:mu0_g_N_pi_params}
\end{table}

\begin{figure}[hp]
\centering\includegraphics[width=\columnwidth,height=58mm]{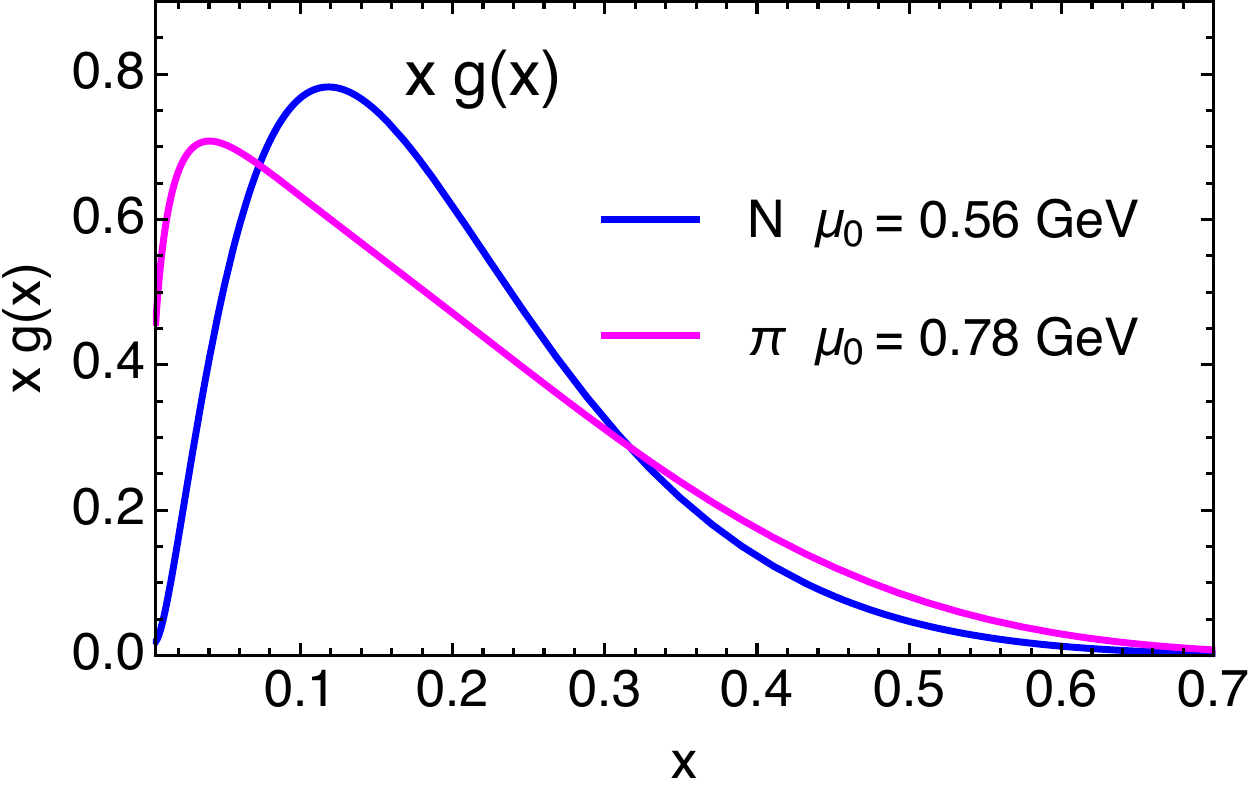}
\caption{The gluon PDFs obtained for the pion (magenta) and the  nucleon (blue) at their respective model scales.      
}
\label{fig:Pi_N_xg_mu0}  
\end{figure}

We employ the  more realistic nucleon description in which  the amplitude introduced in \Eq{eq:Gammax_N} has the form \mbox{$ {\mathcal M}_f(p,P) = D_f(p_c) \, A(p,P) $}
where $A(p,P)$ is a Dirac scalar amplitude. 
This adopted form implements a number of realistic features including momentum dependence for quark number and polarization densities.  
We incorporate properties of the 3-quark description of  a spin up proton that come from a generalization of the $SU(6)$ spin and isospin state~\cite{Close:1979bt,Bhaduri:1988gc}.  That is
\begin{align}
| \Psi \rangle = \frac{\mathcal A}{\sqrt{2}}\,\Big( |\chi_0\rangle\; |\phi_0\rangle\, |F_0\rangle + |\chi_1\rangle\; |\phi_1\rangle\, |F_1\rangle \Big)\, | \chi_c\rangle ~,
\label{eq:Gen_Psi_state}
\end{align}
where $\chi_c$ is the (antisymmetric) color singlet state, $\chi_s(1;2,3)$ is a 3-quark Pauli spin state with a pair coupled to spin $s$ and that coupled to the third quark to make $S_z = 1/2$, and the $\phi_i$ are the corresponding isospin states.  The  $F_\alpha $ are  symmetric spatial states \mbox{$F_\alpha=f_\alpha(p_1)\,f_\alpha(p_2)\,f_\alpha(p_3)$}, with \mbox{$\alpha= 0,1$} and \mbox{${\mathcal A} = 1 - E_{12} - E_{13}$} imposes overall antisymmetry.  The standard SU(6) state corresponds to  \mbox{$f_0 = f_1$} in which case $| \Psi \rangle$ is automatically antisymmetric without the need for operator ${\mathcal A}$. 
After normalization of ${\mathcal M}_f(p,P)$ to quark numbers $n_f$, the resulting quark polarization densities ${\mathcal P}_f(p)$ are such that in the SU(6) limit the standard result  \mbox{$( {\mathcal P}_u,{\mathcal P}_d) = (4/3, -1/3)$} is recovered.  The expression for $A(p,P)$ in terms of the $f_\alpha(p)$ is given in the Appendix.

\begin{table}[tbp]
\ra{0.9}

\begin{tabular}{c|c|cccc}\hline
                  & $\mu~(GeV)$  & \hspace{1mm} $\langle x \rangle_{u_{\rm v}} $ \hspace{1mm}  & \hspace{1mm} $\langle x \rangle_{d_{\rm v}} $ \hspace{1mm} &  \hspace{1mm} $ \langle x \rangle_{\rm sea} $  &   \hspace{1mm}$\langle x \rangle_{\rm g} $ \hspace{1mm}    \\
\hline
\rule{0em}{3ex}    
    N: \hspace*{3mm} Here  \hspace*{4mm}       & 0.56    & 0.443  & 0.187       &  0.171  &  0.199          \\
\hline
\rule{0em}{3ex}                                         
    N: \hspace*{3mm} Here  \hspace*{4mm}       & $\sqrt{5}$   & 0.265  & 0.113     &  0.192  &  0.430     \\
 NNPDF3.0,~\Ref{Ball:2014uwa}    & $\sqrt{5}$   & 0.273  &  0.111      &  0.175  &  0.441            \\
\end{tabular}
\begin{tabular}{c|c|ccccc}\hline 
                   & $\mu_0$~(GeV)  & \hspace{2mm} $\Delta u_{\rm v} $ \hspace{1mm} &  $ \Delta d_{\rm v} $\hspace{1mm} & \hspace{2mm} $\Delta \Sigma_{\rm v} $ \hspace{1mm}  & \hspace{1mm} $2 \Delta G $ \hspace{1mm}   & \hspace{1mm} $g_{\rm A}^{\rm v}$   \hspace{1mm}    \\
\hline
\rule{0em}{3ex}    
       N: \hspace*{3mm} Here  \hspace*{4mm}     &  0.56  & 0.691 & -0.173 &  0.518    &   0.123     &  0.864      \\
\hline 
\end{tabular}
\caption{Momentum fractions and helicities of the nucleon PDFs at the model scale and at a higher scale compared to the indicated global data analysis.  }
\label{N_m1_mu0_mu}
\end{table}

\medskip
\noindent\textbf{Results for Light-cone Momenta and Helicities}:
Models with parameters  set first by reproduction of scale-independent observables such as hadron masses and decays, do not have a naturally identified resolving scale $\mu_0$ associated with the intrinsic PDFs.  That scale can be determined by what is required to fit, by DGLAP evolution upward, one or more empirical PDF $\langle x \rangle$ from global data analysis.    The infinite subset of diagrams in a Rainbow-Ladder truncation has limited ability to accommodate parton splitting and recombination processes that increase with resolving scale.  The resulting DSE-RL 
$\mu_0$ will be greater than $\Lambda_{\rm QCD}$ and should be less than the QCD factorization scale used to factor cross sections into a perturbative scattering mechanism and PDFs containing all non-perturbative physics at lower scales. 

We use the pion approach from \Ref{Bednar:2018mtf}  to set the infrared strength of the gluon-in-quark (hadron independent) kernel ${\mathcal K}_g(q^2)$ of \Eq{eq:CutG_Kernel} so that  under NLO DGLAP scale evolution~\cite{Altarelli:1977zs} $\langle x \rangle_{\rm g} $ reproduces the JAM global analysis~\cite{Barry:2018ort} at \mbox{$\mu=1.3$}~GeV.   The pion model scale \mbox{$\mu_0 = 0.78$}~GeV was previously determined by $q_{\rm v}(x)$ under the non-singlet version of this evolution~\cite{Bednar:2018mtf}.  In the present singlet evolution case we require both $\langle x \rangle_{\rm g} $ and  $\langle x \rangle_{\rm sea}$  be as close as possible to the JAM values.  In some models~\cite{Ding:2019lwe}, it has often been assumed that suitable conditions at $\mu_0$ are  \mbox{$g(x) = q_{\rm sea}(x)=0 $}  and \mbox{$  \langle x \rangle_{q_{\rm v}} = 1$}.    As emphasized by \Refs{Gluck:1989ze,Diehl:2019fsz} the choice of realistic DGLAP starting conditions at $\mu_0$ (especially a non-zero gluon PDF) adds significant benefit to the quality of PDFs at \mbox{$\mu > \mu_0$}.   Here the reduced $\langle x \rangle_{q_{\rm v}}^\pi(\mu_0)$ due to the established quark-in-quark effect prevents adoption of the minimal boundary condition  \mbox{$\langle x \rangle_{\rm sea}^\pi(\mu_0)=0$}; it would require taking \mbox{$\langle x \rangle_{\rm g}^\pi(\mu_0) = 0.354$}  which is untenable because it is already greater than the JAM value at \mbox{$\mu=1.3$}~GeV and will only increase on evolution.   The present model scale momentum fractions that minimize the RMS deviation from JAM values at \mbox{$\mu=1.3$}~GeV are displayed  in \Table{Pion_m1_mu0_mu} and the associated  strength parameter of ${\mathcal K}_g(q^2)$ is shown in \Table{Params}.   The present $2\langle x \rangle_{u_{\rm v}}^\pi(\mu_0)$  is identical to earlier recent work~\cite{Bednar:2018mtf} that also recognized the momentum carried by the gluon-in-quark effect.  
\begin{figure}[tbp]
\centering\includegraphics[width=\columnwidth,height=58mm]{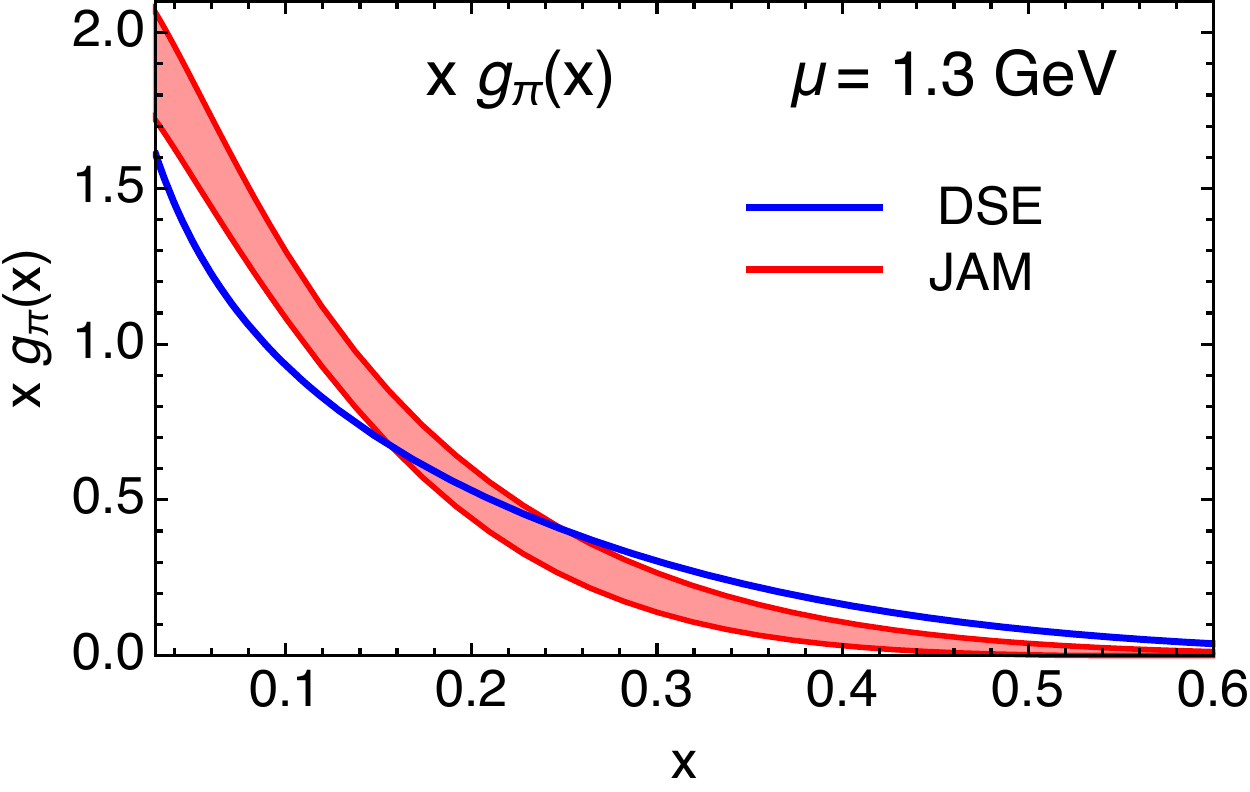}\\[0.5em]
\centering\includegraphics[width=\columnwidth,height=58mm]{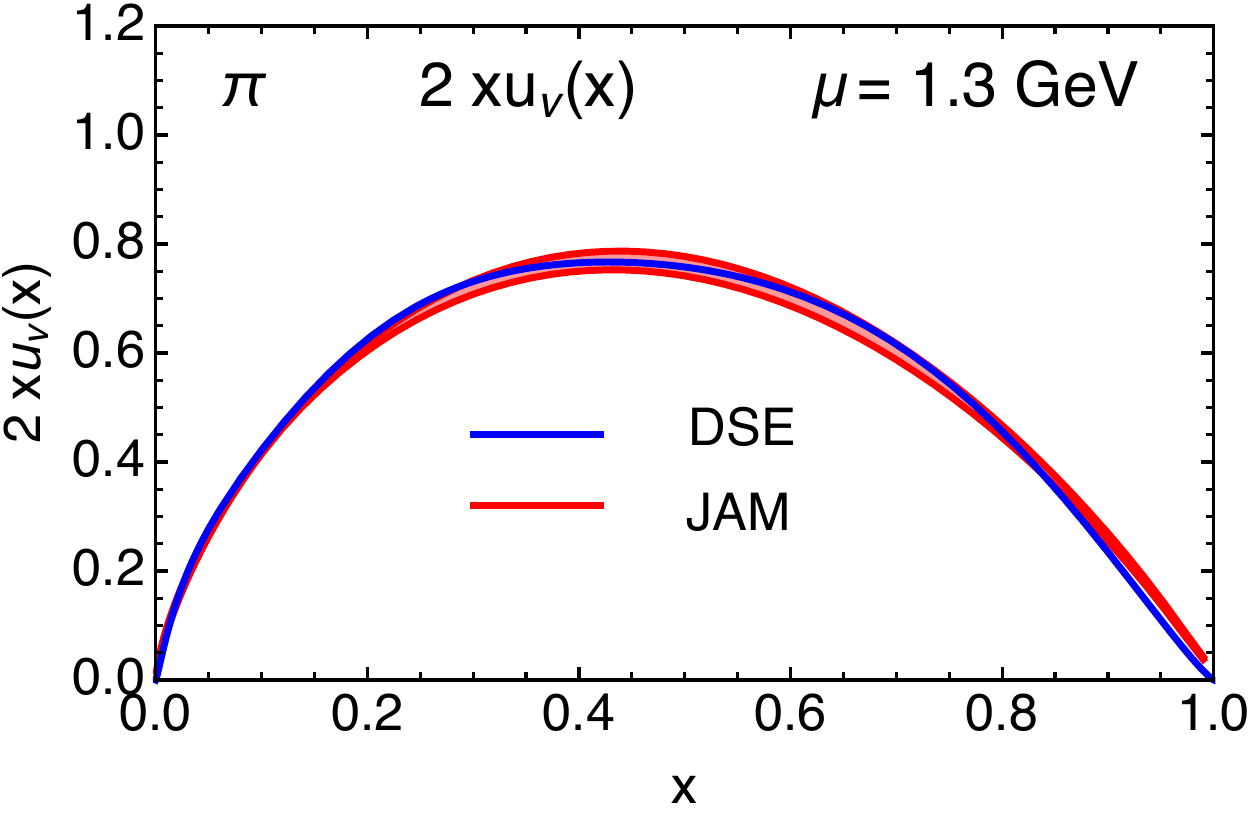}
\caption{{\it Top panel:} PDF of the gluon dressing within the valence quarks of the pion at scale $1.3$~GeV.  Blue solid curve is the present DSE calculation;  the  red band is the result of the 2018 JAM  global data analysis~\protect\cite{Barry:2018ort} with associated uncertainty indicated. 
{\it Bottom panel:} PDF of the valence quarks of the pion compared to the JAM analysis  at scale $1.3$~GeV.
}
\label{fig:Pi_gx_qv_13JAM}  
\end{figure}

With ${\mathcal K}_g(q^2)$ thus set, the employed nucleon amplitude then allows calculation of $\langle x \rangle_{q_{\rm v}}^{\rm N}(\mu_0)$ and $\langle x \rangle_g^{\rm N}(\mu_0)$, while $\langle x \rangle_{\rm sea}^{\rm N}(\mu_0)$ is obtained from the sum rule.  NLO evolution up to \mbox{$\mu=\sqrt{5}~{\rm GeV}$}  and comparison with the NNPDF3.0 analysis~\cite{Ball:2014uwa} then identifies the nucleon model scale \mbox{$\mu_0 = 0.56 $}~GeV, and results are shown in \Table{N_m1_mu0_mu}.    The present $\langle x \rangle_{u_{\rm v}+d_{\rm v}}^{\rm N}(\mu_0)$ value is necessarily quite smaller than previous work~\cite{Bednar:2018htv} which ignored the gluon-in-quark effect by using the convenience of the Ward Identity vertex Ansatz.   This vertex Ansatz is correct only for the lowest Mellin moment (quark number) of quark PDFs; the 1-loop analysis for a quark target discussed earlier provides a simple illustration.  After NLO DGLAP evolution of the unpolarized $n=2$ Mellin  (momentum) moments to compare with data analysis, the results are shown in \Table{N_m1_mu0_mu}.  The RMS deviation of each set of 3 moments from the data analysis is typically 0.03 or less in each case.  

In the lower part of \Table{N_m1_mu0_mu} we display the results for quark and gluon helicities. The valence quark helicity portion $\Delta \Sigma_{\rm v}$ of  the nucleon spin is similar to recent LQCD results~\cite{Alexandrou:2017oeh}. The complete quark helicity, \mbox{$\Delta \Sigma = \Delta u^{+} + \Delta d^{+} $} with  \mbox{$q^{+} =$} \mbox{$q + \bar q$},  requires the sea contribution which we do not produce in the present simple model.  If instead the sea helicity is taken from the polarized PDF analysis of \Ref{deFlorian:2009vb}, then  \mbox{$2 (\Delta \bar u + \Delta \bar d) \sim - 0.15 $} at a scale of 1~GeV.   Together with the present valence result, this indicates \mbox{$\Delta \Sigma  \sim 0.368 $}, a value close to global PDF analyses~\cite{deFlorian:2009vb,Ethier:2017zbq}.  The differing  scales of the components  suggests caution, but the indications are promising.  The gluon helicity portion of nucleon spin, $2 \Delta G$ obtained here at the low model scale, is comparable to the value $0.166$ obtained by a recent global polarized PDF analysis at starting scale 1~GeV~\cite{Leader:2014uua}.    
A more comprehensive treatment of nucleon spin including quark and gluon orbital angular momentum is under investigation.

The valence isovector axial charge \mbox{$g_{\rm A}^{\rm v}(\mu_0) = \Delta u_{\rm v} - \Delta d_{\rm v} $} is not strictly scale invariant, unlike the physical \mbox{$g_{\rm A} = \Delta u^{+} - \Delta d^{+}$}.   Under the common assumption of a flavor symmetric sea, \Table{N_m1_mu0_mu} would indicate  \mbox{$g_{\rm A} \approx g_{\rm A}^{\rm v} \approx$} \mbox{$0.864 $}, significantly below the experimental value $1.26$.   It is typical for a relativistic model of dressed valence quarks to produce \mbox{$g_{\rm A}^{\rm v} \leq 1.0$}, see e.g. \Ref{Eichmann:2011pv}.  However global analysis of polarized PDFs~\cite{deFlorian:2009vb} indicates a sea contribution  \mbox{$2 (\Delta \bar u - \Delta \bar d) \sim 0.3 $} at a scale of 1~GeV; this addition to the present valence  calculation yields 1.16 for $g_A$.  Again despite the different scales involved, this indication is promising.   A treatment  of the Dirac, flavor and momentum structure of the nucleon amplitude that improves upon the present simple illustrative model could  improve all helicity related quantities and such effects are under investigation.     
%
\begin{figure}[tbp]
\centering\includegraphics[width=\columnwidth,height=58mm]{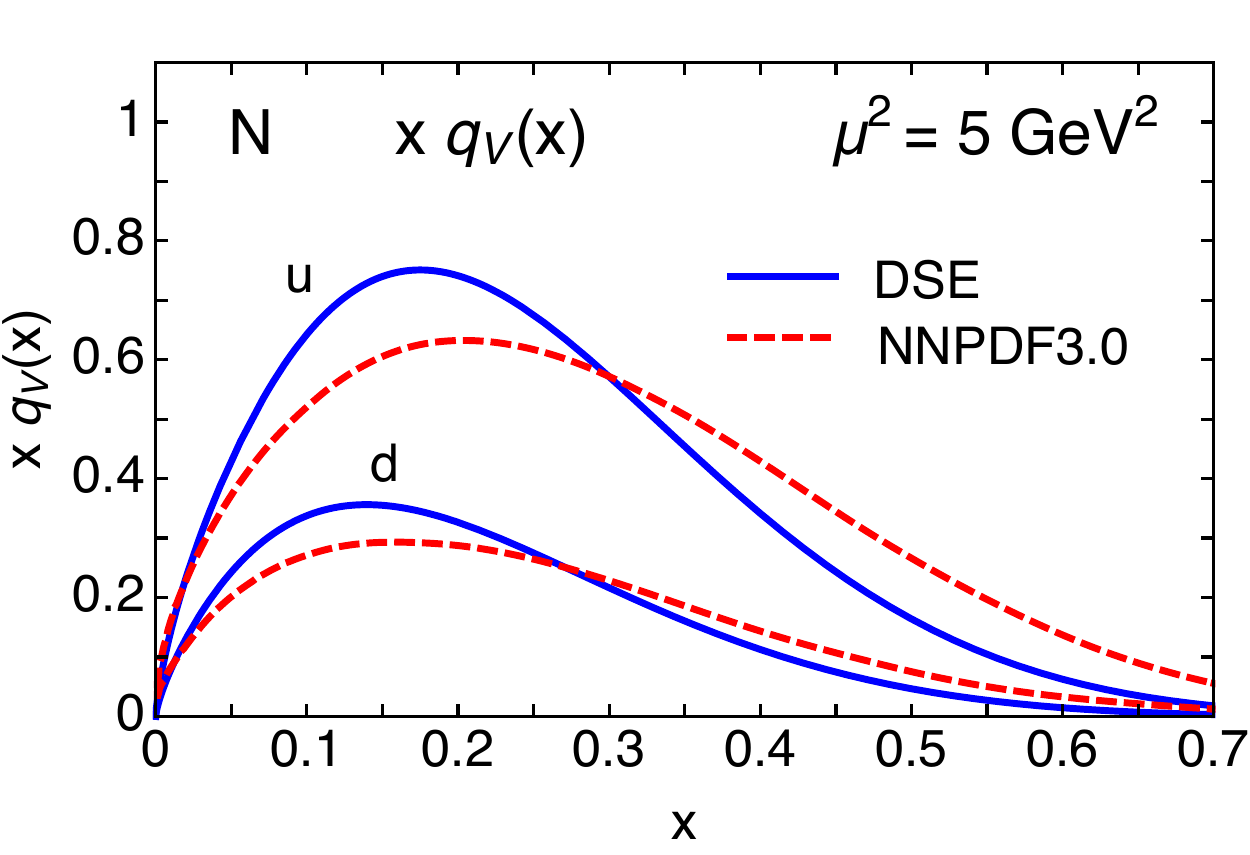}\\[0.5em]
\centering\includegraphics[width=\columnwidth,height=58mm]{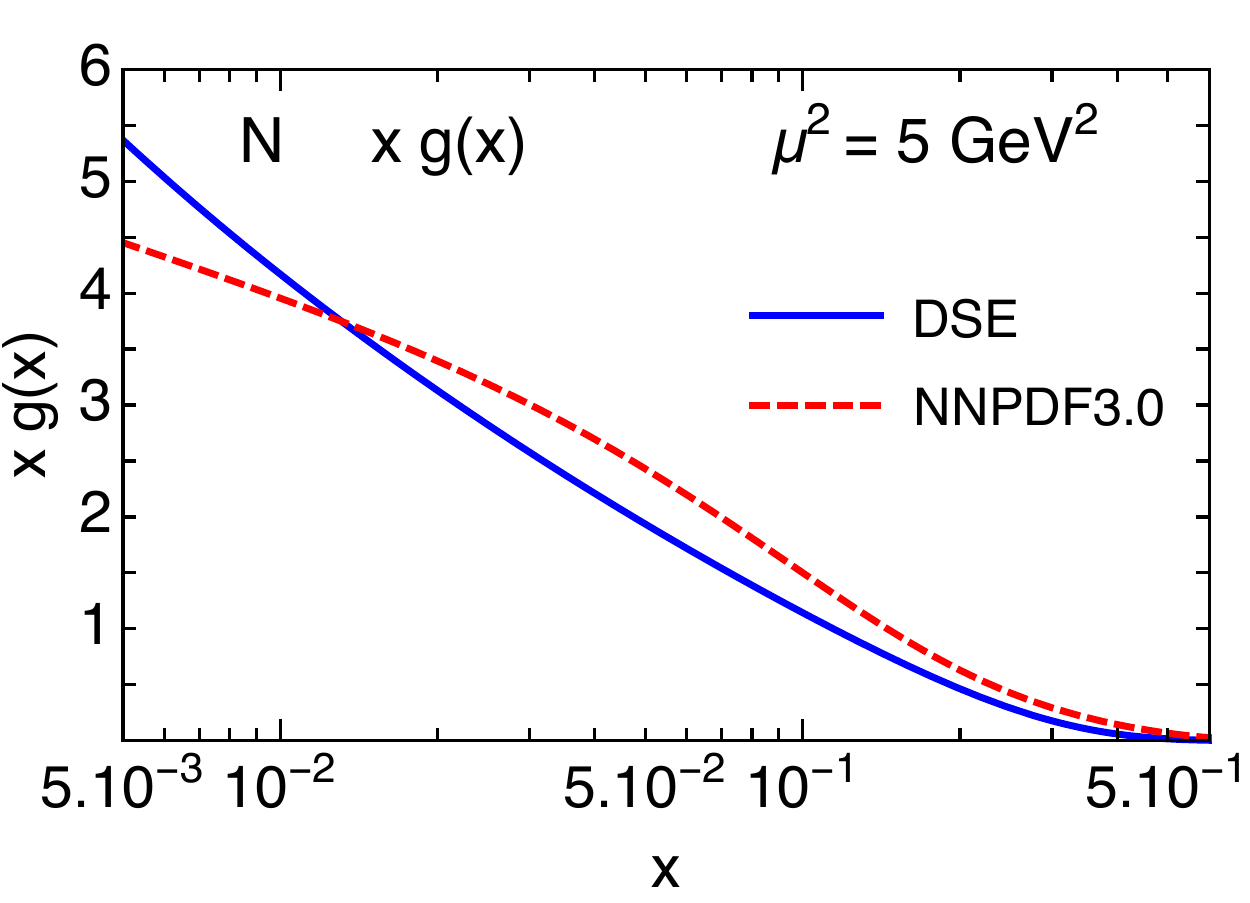}
\caption{{\it Top panel:}  PDFs of the nucleon's valence quarks at scale $\mu^2 = 5$~GeV$^2$.    Blue curve is from the present work and the red dashed curve is from  the NNPDF3.0 data analysis~\protect\cite{Ball:2014uwa}.
{\it Bottom panel:} The  PDF of the gluon-in-quark of the nucleon compared to the NNPDF3.0 analysis result.
}
\label{fig:N_udvx_Gx_5GeV2}  
\end{figure}

The DSE-RL approach here compares well with the present LQCD consensus for $\langle x \rangle_g^{\rm N}$ at 2~GeV~\cite{Deka:2013zha,Yang:2018bft,Alexandrou:2017oeh,Yang:2018nqn,Fan:2020cpa} as follows: 
\begin{align}
\begin{tabular}{c||c|c|c|c|c|c} 
  Here     &   \Ref{Deka:2013zha}  &  \Ref{Yang:2018nqn}    &  \Ref{Alexandrou:2017oeh}  &   \Ref{Yang:2018bft} &  \Ref{Fan:2020cpa} &     ave      \\               
\hline
            0.42             &     0.33       &   0.42         &       0.27          &    0.47    &    0.41            &      0.38       \\
\end{tabular}~.
\label{eq:Gm1_LQCD_results}
\end{align}
For the pion at scale 2~GeV, we obtain \mbox{$\langle x \rangle_g^\pi = 0.343$}. 

\medskip
\noindent\textbf{Results for Quark and Gluon PDFs}:
At $\mu_0$ the PDF moments for \mbox{$m < 6$} were obtained by numerical integration.   
For larger $m$ numerical treatment of the integration over $p$ in \Eq{eq:gen_q_moms_N} and \Eq{eq:gen_g_moms_N}  is  challenging because the vertices $\Gamma_{q/g}^{(m)}(p)$ contain a factor $(p\cdot n/P\cdot n)^m$.   Nevertheless, the integrals are analytically convergent for arbitrarily large $m$ because the result can only depend on scalar products of pairs of the external 4-vectors ($P, n, \nu$, where $\nu$ is polarization).  Since $n\cdot n = 0$ the integrand factor  $(p\cdot n)^m$ has to partner with a factor $(p\cdot P)^{m^\prime}$ with $m^\prime \geq m$ which has to be generated by  the integrand term $\mathcal M_f(p,P)$ as seen via a Taylor expansion of it about \mbox{$p\cdot P =0$}.  The coefficients involve corresponding high order derivatives of  $\mathcal M_f(p,P)$ and each  increases the power of the $p$-dependent  denominator.   Thus the integral remains explicitly convergent with increasing $m$, with its domain of support shifting steadily toward the UV such that for $m > 6 $ only the ranges and power law fall-off of the ingredient factors are seen to be relevant.  The integral can then be cast into standard Feynman representation for which the results are known in algebraic form.   This is done by fitting all elements including vertex amplitudes and propagators to quadratic form denominators respecting the ranges and power law indices.

We use moments up to $m = 30$ to clearly observe the asymptotic behavior $c/m^{h+1}$ and so identify the end point behavior $(1-x)^h$.  This produced \mbox{$h_q^\pi=2$} and \mbox{$h_g^\pi=3$} for the pion and 
\mbox{$h_q^{\rm N}=3$} and \mbox{$h_g^{\rm N}=4$} for the nucleon.  These exponents reflect the lowest non-zero derivative at the end point $x=1$ which in turn reflects the  UV limit \mbox{$Q^2 \to \infty $} implicit in the derivation of the asymptotically hard scattering quark counting rules and the Drell-Yan West relation~\cite{Drell:1969km,West:1970av}.   The gluon end point exponents are consistent with the gluon being sub-leading to its quark source. 

To obtain the $x$ dependence at model scale the moments are fit to the moments of \mbox{$f(x) = $} \mbox{$N\, x^\beta \, (1-x)^h \,B_4(y)$}
where  \mbox{$y=2 \sqrt{x} -x$} and $B_4(y)$ is the polynomial of degree $4$ that uses the Bernstein basis rather than $y^n$, namely
\begin{align}
B_4(y) =  (1-y)^4 & +a\, 4y(1-y)^3 + b\,6y^2(1-y)^2 \nonumber \\
&+ c\,4y^3(1-y) +d\,y^4 ~.
\label{eq:BernsteinB4}
\end{align}
This choice has proved quite efficient in global PDF analyses~\cite{Dulat:2015mca} because it significantly reduces overlaps in the $x$ domains influenced by the parameters $a,b,c,d$.
The obtained values of the parameters  for the model scale gluon PDFs are displayed in \Table{tab:mu0_g_N_pi_params}.  The resulting $x$ dependence is displayed in \Fig{fig:Pi_N_xg_mu0}.
After NLO DGLAP evolution~\footnote{The present exploratory model is not designed to produce the quark sea at model scale, so we employ the data analysis results from \Ref{Gluck:1999xe} for the pion at \mbox{$\mu_0 = 0.632$}~GeV, and from \Ref{Gluck:2007ck} for the nucleon at \mbox{$\mu_0 = 0.707$}~GeV and matched to the present $\langle x \rangle_{\rm sea}$ values.} the resulting pion PDFs are displayed in  \Fig{fig:Pi_gx_qv_13JAM} at scale \mbox{$\mu = 1.3~{\rm GeV}$} and compared with the results of the JAM global analysis~\cite{Barry:2018ort}.  

The same procedures are applied to the nucleon, and after NLO DGLAP scale evolution to $\mu = \sqrt{5}~{\rm GeV}$,   the PDFs are displayed in \Fig{fig:N_udvx_Gx_5GeV2} in comparison with the data analysis results from NNPDF3.0~\cite{Ball:2014uwa}.    The quark sector relates to experiment somewhat better than the earlier work~\cite{Bednar:2018htv} that employed a Faddeev equation description of the nucleon but did not account for the interrelated quark-in-quark and gluon-in-quark effects.   A recent lattice calculation obtains isovector momentum \mbox{$\langle x \rangle_{\rm u_v-d_v} = 0.16$} at 2~GeV~\cite{Detmold:2020snb}.  Our present result compares well to this and to the value  0.162 from the NNPDF3.0 analysis.

\medskip
\noindent\textbf{Summary and Outlook}:
We extend the DSE-RL approach to enable the calculation of the gluon PDF attributable to the dressing of quarks.  Due to the strength of dynamical chiral symmetry breaking, this quark dressing mechanism is expected to produce most of the gluon PDF of light-quark hadrons.  We obtain the interrelated quark and gluon parton momentum fractions using an approach based on the infinite subset of diagrams implemented by the  Rainbow-Ladder truncation of the Dyson-Schwinger equations applied to the pion and nucleon at their natural model scale.  We find the dressing gluon carries about 20\% of the lightcone momentum fraction for both hadrons.    From calculated moments $\langle x^m \rangle_g$ up to \mbox{$m=30$} we identify the dressing gluon-in-quark $g(x)$ in the pion and nucleon.   To enable NLO DGLAP evolution to higher scales to compare with existing data analysis, we employ the valence $q_{\rm v}^\pi(x)$ produced in previous work within this approach and calculate $q_{\rm v}^{\rm N}(x)$ within the present exploratory nucleon model.     The high $x$ end point behaviors of $g(x)$ are found to be $(1-x)^h$ with $h_{\pi}=3$ and $h_{\rm N}=4$; as expected on physical grounds these are 1 greater than the exponents of the corresponding $q(x)$ which are the sources.  

For this first exploration of the gluon-in-quark PDF, we have used the triangle diagram in Landau gauge for $q_{\rm v}(x)$ and thus have ignored the Wilson line contribution.   As an estimate of its magnitude, we have tested a variation of  the model scale boundary condition to start the upward evolution of the $n=2$ Mellin moments for the pion.  Use of $\langle x \rangle_{q{\rm v}}^\pi+\delta_{\rm W}$ and 
$\langle x \rangle_{q{\rm s}}^\pi-\delta_{\rm W}$, with the gauge-invariant $\langle x^m \rangle_g^\pi$ fixed, shows that the new minimized RMS deviation from JAM momenta at 1.3~GeV can lower the previous 0.025-0.03 to 0.01 when $\delta_{\rm W} \sim -0.035$, with a slight increase of the favored model scale from 0.78 to 0.8~GeV.   This suggests that the Wilson line effect is about 3.5\% for the lightcone momentum;  this is much less important that other issues that need to be addressed.

The present results for the pion add $\langle x^m \rangle_g$ and  $g(x)$ to the previously published quark results~\cite{Bednar:2018mtf} of this parton-in-parton approach.   The dynamics of gluon exchange between different valence quarks is found to be down by a factor of $50$ or more in its contribution to $\langle x \rangle_g$; a future work will document this.  The results here for the nucleon are new for  all elements.   The simple model for the nucleon amplitude used here for exploration produces results that are consistent with LQCD and experiment for unpolarized PDFs but are deficient in certain respects for gluon helicity.   This is likely due to the simplicity of the presently employed modified SU(6) model nucleon amplitude.   A generalized amplitude is under study.  Improved QCD-based studies of the gluon PDF within hadrons will help prepare for experimental results from the anticipated Electron-Ion Collider~\cite{Aguilar:2019teb}.

\medskip
\noindent\textbf{Appendix: Form of interaction kernels}:
The interaction kernels in \Eq{eq:RL_Kernel} and  \Eq{eq:CutG_Kernel}, which generate the quark vertices associated with the quark and gluon PDFs respectively,  employ 
\begin{align}
\tilde{\alpha}_s(q^2) = \pi\, \gamma_m \; \left[ \frac{1}{2} \ln \Big( \tau +
        \left(1 + q^2/\Lambda_{\rm QCD}^2\right)^2 \Big) \right]^{-1}   \, ,
\label{eq:tilde_alpha}        
\end{align}
which extrapolates  to the 1-loop coupling $ \alpha_s(q^2)$ in the ultraviolet.  The second interaction kernel also employs
\begin{align}
\tilde{Z}_3(q^2)=   
         \left[ \frac{1}{2} \ln \Big( \tau +  \left(1 + q^2/\Lambda_{\rm QCD}^2 \right)^2 \Big)  \right]^{-\frac{23}{50}}    \, ,
\label{eq:tilde_Z3}        
\end{align}
which extrapolates to the ultraviolet $q^2$ behavior of the 1-loop Landau gauge renormalization quantity $Z_3(q^2,\Lambda^2)$.  We use \mbox{$\Lambda_{\rm QCD}= 0.234 $}~GeV.   Apart from the fixed quantities \mbox{$\tau={\rm e}^2-1$} and  \mbox{$N_f=4$}, the parameters are  given in \Table{Params}.   

\begin{table}[t!]
\ra{0.9}

\begin{tabular}{c|cc|cc}\hline 
              &\hspace{2mm} $D_{\rm RL} $ \hspace{1mm} &  $\omega$\hspace{1mm} & \hspace{1mm} $ D{\rm gg}$ \hspace{1mm} &  \hspace{1mm} $ \omega_g$ \hspace{1mm}        \\
\hline
\rule{0em}{3ex}    
 ${\mathcal K}_{\rm RL}(q^2)\,, {\mathcal K}_g(q^2)$ \hspace{1mm} &  37.324 & 0.5 &  2.98 & 0.53     \\
\hline
\rule{0em}{3ex}     
             &\hspace{2mm}  $ M_{\rm D} $ \hspace{1mm}    & \hspace{1mm} $ N_1/N_0 $ \hspace{2mm} & \hspace{1mm} $ R_{\rm s}$   \hspace{1mm} &  \hspace{1mm} $ R_{\rm v}$  \hspace{1mm}       \\      
\hline
\rule{0em}{3ex}      
  $A_N(p,P)$  &    1.005  &       -1.28      &     0.5       &      0.8               \\
\hline
\end{tabular}
\caption{ Parameters to represent the interaction kernels of \Eq{eq:RL_Kernel} and \Eq{eq:CutG_Kernel} and the model nucleon amplitude of \Eq{eq:Nucl_Ampl}.   }
\label{Params}
\end{table}

\medskip
\noindent\textbf{Appendix: The Nucleon Model:} 
For the amplitude $A(p,P)$ we employ the form 
\begin{align}
A(p,P) = N \,\frac{ \small\sum_{\alpha\prime \alpha} \, f_{\alpha^\prime}(p)\, f_\alpha(p)}{ (K^2 + M_{\rm D}^2) }~,
\label{eq:Nucl_Ampl}
\end{align}
with the functions associated with the spin$-1$ and spin$-0$ terms of the underlying nucleon state in \Eq{eq:Gen_Psi_state} having the form \mbox{$f_\alpha(p) = N_\alpha/( (p - P/3 )^2 + R^2_\alpha ) $}. 
Note that $p - P/3$ is the relative momentum of the active quark and  spectator system, while the latter has momentum \mbox{$K = P - p$}.   The ratio $N_1/N_0 $ replicates the relative infrared strength of the spin-1 and spin-0 $qq$ correlations within the Faddeev amplitudes employed  in \Ref{Bednar:2018htv}, while $N$ is determined by valence quark number.  The parameters are displayed in \Table{Params}.  

\smallskip
\noindent\textit{Acknowledgments:}
We acknowledge beneficial discussions  with  Chao Shi and Anthony Thomas.  We appreciate the information and data analysis results provided by Patrick Barry and the JAM Collaboration.  This work was supported by the National Science Foundation, grant no.\ NSF-PHY1516138, and by the U.S. Department of Energy, Office of Science, Office of Nuclear Physics, contract no. DE-AC02-06CH11357 and contract no. DE-FG02-97ER41014.


%

\end{document}